# MITIGATION AND CONTROL OF INSTABILITIES IN DAFNE POSITRON RING*

Alessandro Drago, David Alesini, Theo Demma, Alessandro Gallo, Susanna Guiducci, Catia Milardi, Pantaleo Raimondi, Mikhail Zobov
Istituto Nazionale di Fisica, Nucleare Laboratori Nazionali di Frascati, Frascati, Italy

*Abstract*

The positron beam in the DAFNE e+/e- collider has always been suffering from strong e-cloud instabilities. In order to cope with them, several approaches have been adopted along the years: flexible and powerful bunch-by-bunch feedback systems, solenoids around the straight sections of the vacuum chamber and, in the last runs, e-cloud clearing electrodes inside the bending and wiggler magnets. Of course classic diagnostics tools have been used to evaluate the effectiveness of the adopted measures and the correct setup of the devices, in order to acquire total beam and bunch-by-bunch currents, to plot in real time synchrotron and betatron instabilities, to verify the vertical beam size enlargement in collision and out of collision. Besides, to evaluate the efficacy of the solenoids and of the clearing electrodes versus the instability speed, the more powerful tools have been the special diagnostics routines making use of the bunch-by-bunch feedback systems to quickly compute the growth rate instabilities and the bunch-by-bunch tune spread in different beam conditions.

## INTRODUCTION

In the DAFNE e+/e- collider [1-2], in operation since 1997, the positron beam has always been suffering from strong instabilities [3] mainly due to parasitic e- clouds. In order to cope with them, several approaches have been adopted along the years: flexible and powerful bunch-by-bunch transverse feedback systems, solenoids around the straight sections of the vacuum chamber and, in the last runs, e-cloud clearing electrodes.

Metallic electrodes have been designed to absorb the photo-electrons in the DAFNE positron ring. They have been inserted in the wiggler and bending magnet vacuum chambers and have been connected to external voltage generators.

The dipole electrodes have a length of 1.4 or 1.6 m depending on the considered arc, while the wiggler ones are 1.4 m long. They have a width of 50 mm and a thickness of 1.5 mm.

The electrodes have been made in copper and have a distance of 0.5 mm from the vacuum pipe. This small distance has been chosen to reduce the beam coupling impedance of the devices. The distance is guaranteed by special ceramic supports made in SHAPAL and distributed along the electrodes.

The distance of the electrodes from the beam axis is 8 mm in the wigglers and 25 mm in the bending magnets.

Analytical calculations and electromagnetic simulations have been done to estimate the power released from the beam to the electrodes. We expect a maximum temperature increase of the order of 100$^{o}$C with a 2A beam for the wiggler electrodes. This temperature increase has been considered acceptable since the electrodes have been heated up to this level without damage and also because it is in the range of operation of all the components (SHAPAL and feedthroughs).

The electrodes are connected to external generators and have been tested (with the beam) applying dc voltages up to 250 V.

RF measurements have been also done to precisely measure the resonant frequencies of the electrodes modes.

RF measurements have been performed before and after the electrode installation by using a network analyzer. We have done two types of measurements: reflection coefficient at the feedthrough port and transmission coefficient between one BPM near to the strip and the feedthrough. In both cases it was possible to measure the resonant frequencies of the stripline modes.

The electrode impedance consists of two contributions: resistive wall impedance due to a finite conductivity of the electrode and stripline impedance due to the gap created between the electrode and the vacuum chamber wall.

The estimated low frequency broadband impedance of the electrode Z/n is about 0.005 Ohm that should give a small contribution to the total ring impedance.

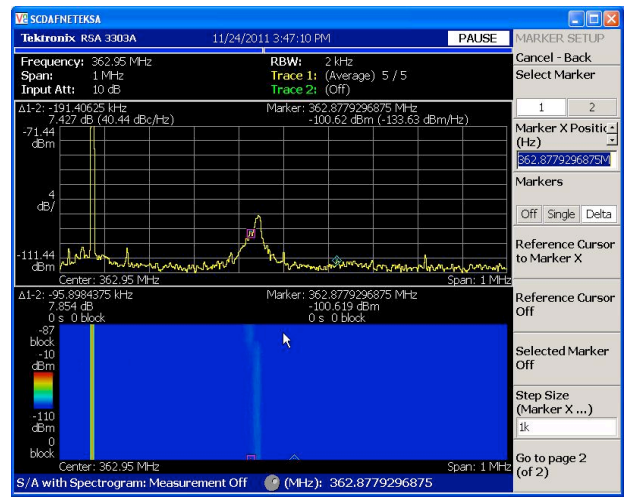

Figure 1: The beam horizontal frequency shift caused by turning off all the 12 electrodes is ~20KHz, that corresponds to a difference in the horizontal tune of ~0.0065.

## MEASUREMENTS ON e+ BEAM

Looking at the effect on the positron beam, measures have been carried on by using a synchrotron light monitor, a FFT spectrum analyzer, and the bunch-by-bunch horizontal and vertical feedback systems.

As shown in Fig. 1, a FFT analyzer (Tektronix RSA3303A) is used to study the beam frequency response from a button pickup. A horizontal frequency shift is evident by turning off all the clearing electrodes. The frequency difference is ~20 KHz corresponding to a difference in the horizontal tune of ~0.0065. Frequency and betatron tune grow up when the electrodes go off. This kind of measure does not separate the signal for each bunch, plotting the response in frequency of the whole beam behavior.

In Fig. 2 a plot from the SLM is shown: turning off progressively the electrodes, a vertical enlargement is evident on the SLM. The beam vertical size goes from 110 to 145 μm. These data are not recorded by a gated camera so they are not a bunch by bunch measure. Furthermore in this measure as well as in the following ones, the positron beam is not colliding with the electron beam.

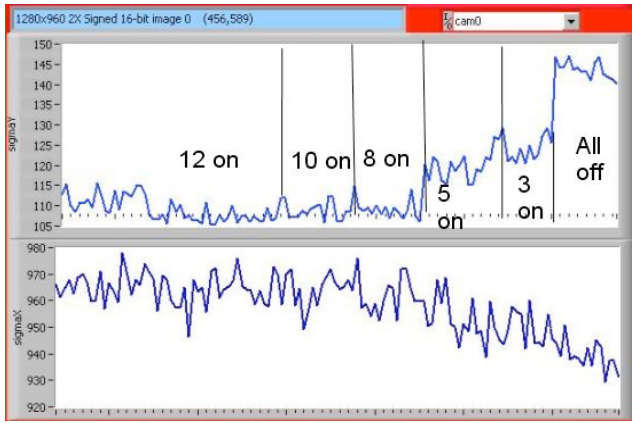

Figure 2: Vertical beam size enlargement due to the clearing electrodes progressive turning off: the beam vertical size goes from 110 to 145 μm.

In DAFNE it is possible to acquire bunch by bunch data using the feedback systems designed to damp the coupled bunch instabilities. The bunch-by-bunch feedback has been developed in 1992-96 by a large collaboration of SLAC, LNF, ALS-Berkeley, [4], [5], and, in the latest upgrade, KEK [6], and it has shown powerful diagnostics capabilities [7-9]. In the first design was implemented only for longitudinal damping, then, in the following upgrades, also as transverse feedback. Recently a new upgrade including 12 bits analogue conversion (it was done previously by 8 bits) has been installed at DAFNE [10] for damping transverse instabilities. The feedback has been used in order to produce horizontal instability growth rate measurements versus beam currents (as shown in Fig. 3) testing the performance of the e-cloud clearing electrodes at different voltages.

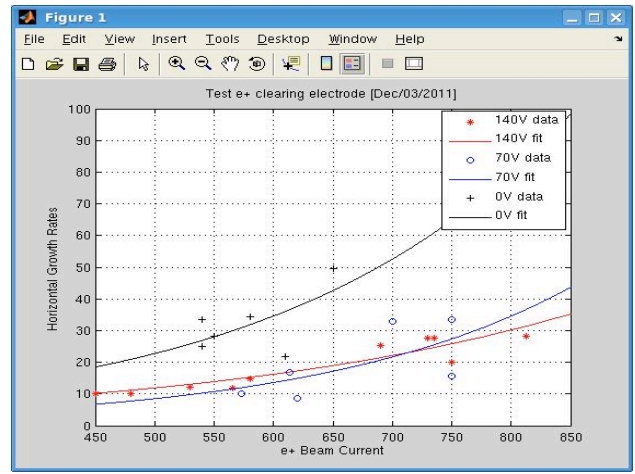

Figure 3: Horizontal instability growth rates (ms-1) versus beam current (mA). Measures done by using the bunch-by-bunch feedback at different voltages applied to the clearing electrodes.

Growth rates (in ms-1) have been measured comparing no voltage with 70V and 140V. In Fig. 3 the effectiveness of the electrodes is evident. After many tests on the reliability with high beam currents, the voltages applied to the electrodes have been increased to 200V (in dipoles) and 250V (in the wiggler magnets).

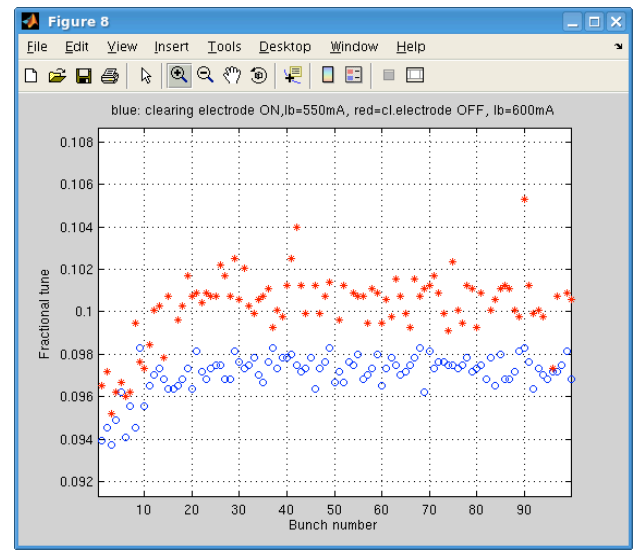

Figure 4: Horizontal fractional tune versus bunch number measured by the bunch-by-bunch feedback system acquiring and averaging 12 M samples of data. Turning off the electrodes in 4 wigglers and 2 dipoles, the horizontal tune goes up.

By using the new version of the feedback based on Xilinx Virtex-5 FPGA, up to 12 Msamples can be stored in the chip memory. After recording the longest tracks, data are downloaded to the server where they are processed. Fractional tune measures versus bunch number are presented in Fig. 4 (horizontal) and in fig. 5 (vertical).

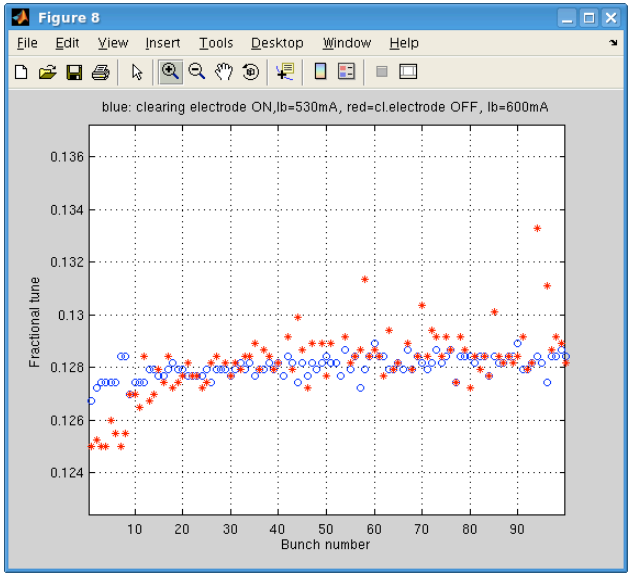

Figure 5: Vertical fractional tune versus bunch number measured by the bunch-by-bunch feedback system. Turning off the electrodes in 4 wigglers and 2 dipoles, the Vertical tune is more scattered.

As it is shown in the Fig. 5, turning off the 4 wiggler electrodes and 2 (over 8) dipole electrodes, the fractional tune goes up by about +0.0040. This value is in very good agreement with the value of +0.0065 (see Fig. 1) achieved tuning off all 12 electrode voltages. It is remarkable to observe that the first 10 bunches in both cases show lower tune values of about 0.02 versus the rest of the bunch train. The case with voltage off shows also a larger tune spread from bunch to bunch.

Comparing these horizontal data with the vertical analysis, first of all it is possible to evaluate the noise of the measure system itself that, looking at the blue circles in the Fig. 5, is in a range < 0.0005 in terms of tune unit. In addition, also from other kind of measures and relative considerations done about the same data acquisition system, the resolution of the acquired data seems correct within ~0.0005 range.

Evaluated the measurement limits, it is remarkable to observe the effect of the wiggler electrodes on the bunch vertical behavior. Indeed both records show a vertical tune around 0.128 but the spread is much smaller (~0.002) with the electrodes on than with the electrodes off (~0.006). So, also if solenoids in the straight regions are considered more useful to damp e-cloud induced vertical instabilities, the wiggler clearing electrodes show interesting positive effects not only in the horizontal plane but also in vertical one.

## CONCLUSION

Metallic clearing electrodes have been inserted in the wiggler and bending magnet vacuum chambers of the DAFNE positron ring to mitigate the instability due to the e-cloud. The electrode placement is complementary to solenoids that are allocated in the straight sections of the e+ ring. Experience with clearing electrodes in the DAFNE positron beam is largely positive: smaller vertical dimensions, less transverse tune spread and slower growth rates clearly indicate a good behavior of these devices. Transverse bunch-by-bunch feedback systems with many diagnostics analysis programs have been proved crucial tools to evaluate solenoid and e-cloud clearing electrode performances. Furthermore, in our opinion, the agreement between the classic instruments and innovative techniques is very good considering experimental measurement errors and finite width of the sidebands.

## ACKNOWLEDGMENT

This work has been partially funded by the HiLumi LHC Design Study. The HiLumi LHC Design Study (a sub-system of HL-LHC) is co-funded by the European Commission within the Framework Programme 7 Capacities Specific Programme, Grant Agreement 284404.